\documentclass[aps,pra,reprint,a4paper,superscriptaddress,floatfix,amsmath,amssymb,amsfonts]{revtex4-1}

\usepackage[utf8]{inputenc}
\usepackage{graphicx}
\usepackage[normalem]{ulem}
\usepackage[usenames,dvipsnames]{color}
\usepackage[nointegrals]{wasysym}
\usepackage{commands}
\usepackage{upgreek}

\usepackage{amsmath}
\usepackage[pdftex]{hyperref}
\hypersetup{pdfauthor={}, bookmarksnumbered=true, pdftitle={}, colorlinks, citecolor=blue, linkcolor=blue, urlcolor=blue}
 
\cleanversion 
\blackwhite 

\begin{document} 


\title{Field effect induced mesoscopic devices in depleted two dimensional electron systems}

\author{N. Bachsoliani}\affiliationMunich
\author{S. Platonov}\affiliationMunich \affiliationPDI
\author{A.D. Wieck}\affiliationBochum
\author{S. Ludwig}\affiliationPDI

\date{\today}


\begin{abstract}
Nanoelectronic devices embedded in the two-dimensional electron system (2DES) of a GaAs/AlGaAs heterostructure enable a large variety of applications from fundamental research to high speed transistors. Electrical circuits are thereby commonly defined by creating barriers for carriers by selective depletion of a pre-existing 2DES. Here we explore an alternative approach: we deplete the 2DES globally by applying a negative voltage to a global top gate and screen the electric field of the top gate only locally using nanoscale gates placed on the wafer surface between the plane of the 2DES and the top gate. Free carriers are located beneath the screen gates and their properties can be controlled by means of geometry and applied voltages. This method promises considerable advantages for the definition of complex circuits by the electric field effect as it allows to reduce the number of gates and simplify gate geometries. Examples are carrier systems with ring topology or large arrays of quantum dots. Here, we present a first exploration of this method pursuing field effect, Hall effect and Aharonov-Bohm measurements to study electrostatic, dynamic and coherent properties.
\end{abstract}

\maketitle 

\section{introduction}

The electric field effect is a powerful tool for nanoelectronics. It is widely used for creating potential barriers in a two-dimensional electron system (2DES) by applying voltages to individual metal gates placed on the wafer surface. When used with multiple individual gates it provides full tunability while being compatible with high mobility wafers. Alternative methods for structuring a 2DES include etching \cite{behringer1987,Lee1989,Ford1990,moon1998} and surface oxidation techniques \cite{held1998,heinzel2001}. While they ensure additional possibilities in combination with in-plane side gates \cite{wieck1990} or metal gates \cite{held1999,platonov2015}, etching and oxidation techniques alone lack tunability. More importantly, they are restricted to wafers with a shallow 2DES causing a close proximity of surface states and doping atoms limiting the  carrier mobility \cite{walukiewicz1984} and the electrostatic stability at the nanoscale (related with the $1/f$ spectrum of charge noise \cite{jung2004,buizert2008,taubert2008}). Applications based on the quantum mechanical coherence of localized carriers require superior control and stability favoring the field effect.  

A straightforward and the most common approach to shape potential landscapes by the field effect, starting from an extended 2DES, is based on the controlled local depletion of the 2DES beneath individual surface gates. This approach works perfectly for relatively small structures with simple topology such as few coupled quantum dots \cite{schroer2007,forster2015} or quantum point contacts. However, an individually tunable one-dimensional array of $N$ quantum dots requires at least $\sim2N$ metal gates, while even more gates are needed for a two-dimensional array or for increased tunability. Failure of a single gate would alter the current path and typically make the entire device useless. Furthermore, non-trivial topologies such as an Aharonov-Bohm ring, allowing carriers to move in a circle around a depleted center, require voltage biasing of a center gate without depleting the surrounding carriers. This has been achieved by implementation of three-dimensional air bridges \cite{buks1998,chang2008}. However, the fabrication of air bridges is rather complex and limited to relatively big structures.

\begin{figure*}[htb]
\begin{center} 
\includegraphics[width=1\linewidth]{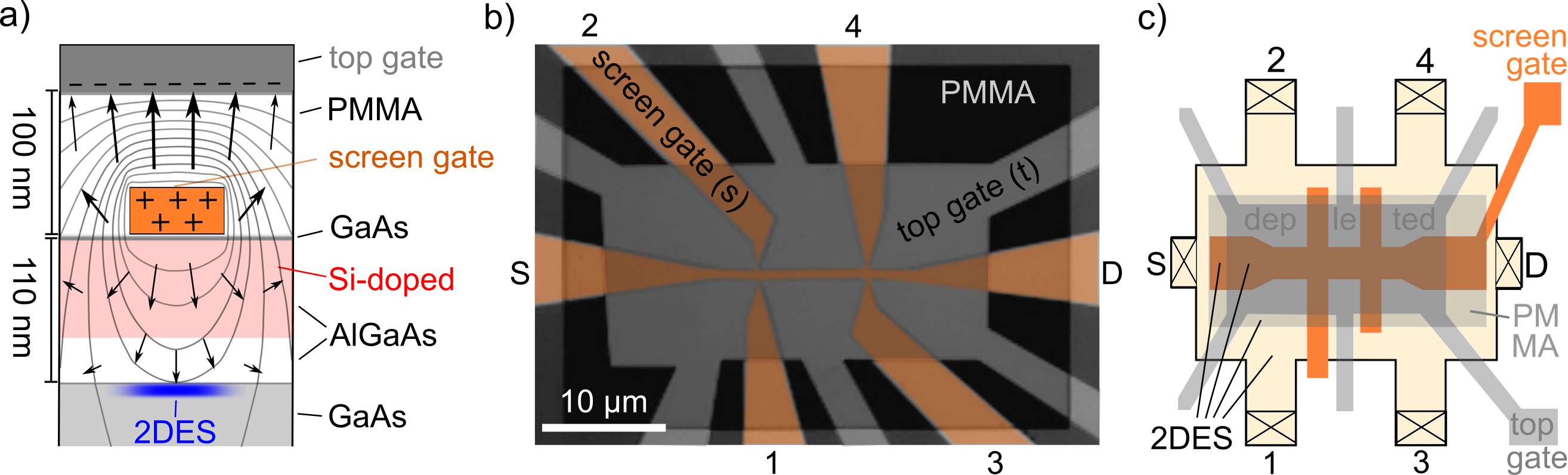}
\caption{a) Sketch of the heterostructure with screen and top gate. The top gate (gray) is biased at $\vt=-1\,$V, the screen gate at $\vs=0.1\,$V and the 2DES (blue) is grounded. Away from the screen gate but below the top gate the 2DES is absent (fully depleted). The electrostatic potential (shown as equally spaced equipotential lines, arrows indicate the field strength) has been calculated by self-consistently solving the Poisson equation using $\epsilon=2$ for cross-linked PMMA and $\epsilon=12.7$ for AlGaAs. For the calculation we considered charges on the gates and in the 2DES but neglected the effect of positively charged doping ions which are immobile at cryogenic temperatures.
b) False-colored optical microscope image of a Hall-bar sample. Orange color indicates the screen gate (s)
\SL{covered by cross-linked PMMA (dark) and finally the top gate (gray).
c) Descriptive sketch of the sample surface. As in panel b) it shows the screen gate (orange) covered by PMMA (light gray, slightly yellow) and the top gate (gray). At the white surrounding the sample surface is etched away such that the 2DES is destroyed. Ohmic contacts to the 2DES for source (S), drain (D) and voltage probes (1, 2, 3, 4) are indicated by crosses. In a Hall-bar measurement, the 2DES is depleted beneath the top gate but not beneath the screen gate. Yellow colored regions are not covered by a gate and always contain conducting 2DES 110\,nm beneath the surface.}
}
\label{fig_1}
\end{center}
\end{figure*}

In this article we propose an alternative method to define complex nanoelectronic circuits based on the field effect, offering full tunability of high quality devices. Compared to common strategies our method simplifies the production of ring topologies and offers the prospect of scalability while limiting the danger of general failure. Our idea, sketched in \fig{fig_1}a,
is based on globally depleting the 2DES using a single top gate while we place nanoscale screen gates between the top gate and the 2DES to locally shield the effect of the top gate and thereby regain free carriers. We electrically isolate the top gate from the screen gates using an about $100\,$nm thick layer of cross-linked PMMA \cite{zailer1996,teh2003}\SL{, while the 2DES at the interface between AlGaAs and GaAs is separate from the screen gates on the wafer surface by another 110\,nm consisting of the following electrically insulating layers:}\scrap{ . In our sample the 2DES at the interface between AlGaAs and GaAs is separated from the surface by 5\,nm of GaAs (capping layer)} \SL{a  5\,nm thick capping layer  of GaAs to prevent oxidation of the surface,}  followed by 70\,nm of homogeneously Si-doped Al$_{0.36}$Ga$_{0.64}$As, and 35\,nm of undoped Al$_{0.36}$Ga$_{0.64}$As \cite{Note1,wieck2000}\footnotetext{Below this, there is a $650\,$nm thick GaAs layer on top of $15$ periods of a superlattice of $5\,$nm Al and $5\,$nm As to stick segregating impurities during the MBE-growth at its interfaces and, thus, to keep the layers above clean. The 2DES is formed in the GaAs-layer close to the interface of the Al$_{0.36}$Ga$_{0.64}$As layer.}. Carrier density and the detailed geometry of the confinement potential depend on the electric field at the 2DES and can be fine tuned by adjusting the \scrap{individual gate} voltages \SL{applied to both, the top gate and the screen gate}. In \fig{fig_1}a we sketch the screening effect on a grounded 2DES for the example of a positively charged screen gate beneath a negatively charged top gate. A global top gate above gates at the surface had been used before for different purposes \scrap{including a}. \SL{In a previous attempt to structure a 2DES, a single top gate was combined with a local dielectric to partially screen the field of the global top gate \cite{Ford1988,Ford1989}. Here, the missing screen gate results in a reduced tunability compared to our approach. A global top gate has also been employed to} decrease telegraph noise \cite{buizert2008} or to incorporate carriers in undoped quantum wells \cite{harrell1999,chen2012}. In the \SL{last two} examples gates on the GaAs surface are used to locally deplete the 2DES while in our case carriers accumulate beneath the screen gate. 

In \fig{fig_1}b we display a scanning electron microscope (SEM) image of an actual Hall-bar sample \SL{and in \fig{fig_1}c a descriptive sketch}. \SL{The screen gate (s) directly on the sample surface is shown in orange, the top gate (t) above, which is electrically isolated by cross-linked PMMA, in light gray.} \scrap{The  top gate (t) is shown in light grey and the screen gate (s) colored orange; the isolating cross-linked PMMA manifests itself as a squared slightly darker region extending across most of the image.} By charging the top gate negatively in respect to the grounded 2DES and a grounded backgate at the bottom of the $540\,\mu$m thick wafer we deplete the 2DES beneath the top gate where\SL{ever} it is not shielded by the screen gate. Below the top gate the shape of the screen gate corresponds to the approximate shape of the 2DES beneath. The screen gate in \fig{fig_1}b defines a Hall-bar with source (S) and drain (D) for current and four side contacts used as voltage probes (1,2,3,4). The top gate includes a large center square and six arms \scrap{which electrically isolate the conducting leads in the periphery of the Hall-bar} \SL{reaching to the outside. The arms have the function to avoid electrical shorts between the six contacts outside of the Hall bar region where they would otherwise be shorted by 2DES (as the yellow areas in \fig{fig_1}c are conducting). }

In our Hall-bar the free carriers are located directly beneath a metal gate, which results in two important differences to traditional devices: the direct vicinity of metal can reduce the disorder potential as charged defects are partly screened by electron rearrangement at the metal surface. At the same time the metal will tend to screen the electron-electron interaction in the 2DES below. In the present article we do not explore this reduced Coulomb interaction but rather demonstrate the general feasibility of our method.

\section{field effect characterization}

For a first characterization of our device
\begin{figure}[htb]
\begin{center}
\includegraphics[width=1\linewidth]{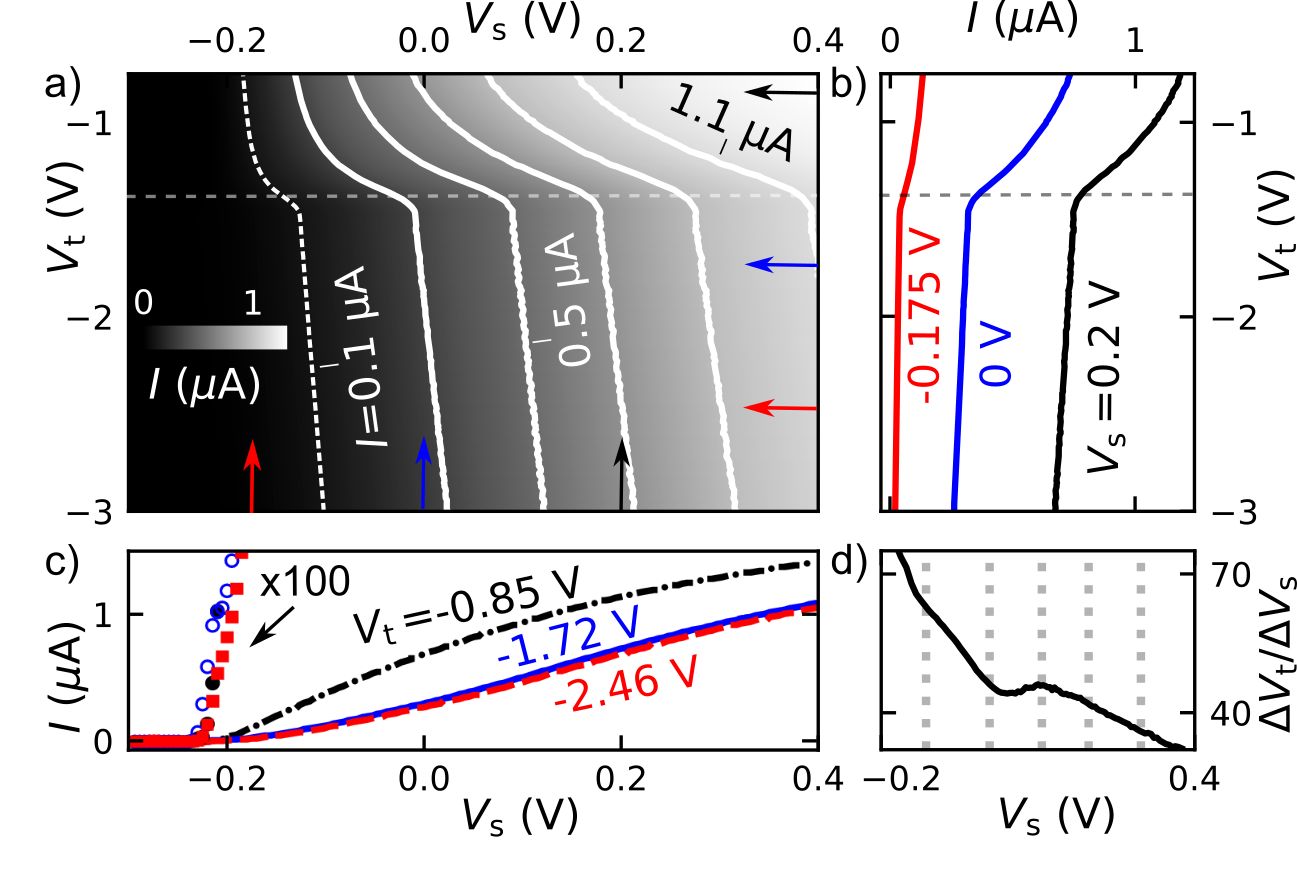}
\caption{a) Current $I$ through the Hall-bar at $\vsd=0.84\,$mV (gray scale and lines of constant current at an interval of $0.2\,\mu$V) as a function of top gate $V_\text{t}$ ($y$-axis) and screen gate \vs\ ($x$-axis) voltages. The horizontal dashed line at $\vt \equiv\vtd= -1.48\,$V indicates the onset of depletion of the 2DES below the top gate away from the screen gate. 
b) and c) Vertical and horizontal cuts $I(\vt)$ and $I(\vs)$ from panel a (fixed voltages \vs\ and \vt, respectively, are indicated by color coded arrows).
\SL{Symbols in panel c) represent identical data multiplied by a factor 100 to demonstarte complete pinch-off for $\vs<-0.228\,$V independent of \vt.}
d) The slope $dV_\text{t}/dV_\text{s}$ of the constant current lines versus $\vs$ at $\vt=-2.5\,$V. Vertical dashed lines indicate intersections with constant current lines in panel a.
 }  
\label{fig_2} 
\end{center}
\end{figure} 
we present in \fig{fig_2}{} the current flowing between source and drain contacts (while the side contacts are left floating) in response to a source-drain voltage of $\vsd=0.84\,$mV as a function of both top gate and screen gate voltages $\vt$ and $\vs$. The lines of constant current display a kink at $\vt\equiv\vtd = -1.48\,$V, marked by a dashed horizontal line in panels a) and b), indicating complete depletion of the 2DES for $\vt<\vtd$. The almost constant slope of \scrap{the lines} \SL{each line} of constant current for $\vt < \vtd$ suggest\SL{, for a given value of \vs,} a constant ratio of the capacitances between the Hall-bar and the two respective gates, $\cs/\ct=d\vt/d\vs$\scrap{, which we plot in \fig{fig_2}d versus \vs}. \SL{To keep the current constant, a change in the screen gate voltage by $\delta\vs$ can be compensated by a shift of the top gate voltage by $\delta\vt=-\delta\vs\cs/\ct$. The value of $\cs/\ct$ quantifies the shielding of the influence of the top gate on the 2DES by the screen gate. It depends on the dielectric constants and geometry of the layers, which influence the electric field originating from the top gate.} The coupling ratio\SL{, which we plot in \fig{fig_2}d versus \vs,} takes the large value of $\cs/\ct\simeq75$ at  $\vs\simeq-200\,$mV near depletion where it indicates an efficient screening of the top gate by the screen gate. The gradual increase to $\cs/\ct\simeq45$ at $\vs\simeq300\,$mV indicates a growing influence of the top gate at more positive \vs. \SL{Because the 2DES shaping the Hall-bar is the only variable component of our system, the observed reduction of the shielding effect, as \vs\ is increased, indicates an increase of the Hall-bar width. Variations in the Hall-bar width, in turn, result in a rearrangement of the confinement potential perpendicular to the Hall-bar edges.} \scrap{It suggests that the Hall-bar becomes wider at more positive \vs\ while the potential profile at the Hall-bar edges steepens.} Consequently, the combination of top- and screen gate voltages can be used to tune the \SL{steepness of the confinement at the} Hall-bar edges \scrap{and} \SL{which influence} the shape and stability of quantum Hall edge states \cite{siddiki2010}. Reliable predictions could be achieved employing a Poison-Schr\"odinger solver such as nextnano$^3$ \cite{birner2007}, while breakdown measurements of the quantum Hall effect would provide an experimental test \cite{siddiki2010}. Both ideas go beyond the scope of the present paper but are topics for the future.

\scrap{From measured ratios of capacitances we can quantify the shielding of the top gate by the screen gate: we therefore determine the ratio $\cs/\ct^0$ with $\ct^0$ being the capacitance between the top gate and the 2DES without screen gate in between.}
\SL{To quantitatively evaluate the shielding, we compare the measured capacitance ratio $\cs/\ct$ with the ratio expected without screening $\cs/\ct^0$, where $\ct^0$ denotes the capacitance between the top gate and the 2DES without the existence of a screen gate.} For a first estimate we compare the \SL{measured} depletion voltages of the respective gates $\cs/\ct^0\simeq\vtd/\vs^\text d\simeq1.48/0.23\simeq6.4$. As a result, we find $\ct^0/\ct=\left(\cs/\ct\right)/\left(\cs/\ct^0\right)\simeq75/6.4\simeq11.7$ at $\vs=-0.23\,$V, i.e.\ the screen gate reduces the coupling of the top gate to the 2DES roughly by one order of magnitude. Clearly, this result depends on the geometry details and the applied voltages. \SL{The accuracy of the above numbers is around 10\% reflecting the accuracy in determining the pinch-off voltages.} 

\SL{In a second approach we compare our first estimate based on direct measurements with the prediction of a simple plate capacitor model, assuming two separate plate capacitors, one between the top gate and the 2DES ---but without screen gate--- and the other between the screen gate and the 2DES. The model predicts $\cs/\ct^0\simeq1+\epsilon_\text{AlGaAs}/d_\text{AlGaAs}\times d_\text{PMMA}/\epsilon_\text{PMMA}$, where the capacitor between the top gate and the 2DES contains two layers of dielectricum, $d_\text{PMMA}=100\,$nm of PMMA and $d_\text{AlGaAs}=110\,$nm of AlGaAs. We determine the required dielectric constant of our cross linked PMMA from our measured depletion voltage $\vtd=-1.48\,$V of the top gate and the carrier density of the 2DES $\ns^0\simeq1.45\times10^{11}\,\text {cm}^{-2}$ at grounded gates, $\vt=\vs=0$, based on Hall measurements. Using our simple plate capacitor model, we find $\ns=\left(d_\text{PMMA}/\epsilon_\text{PMMA}+d_\text{AlGaAs}/\epsilon_\text{AlGaAs}\right)\vt^d/\epsilon_0$ with $\epsilon_0$ being the vacuum permeability. Using the literature value $\epsilon_\text{AlGaAs}=12.7$ \cite{Samara1983} we find $\epsilon_\text{PMMA}\simeq 2.0$. Finally, our plate capacitor model predicts  $\cs/\ct^0\simeq1+\epsilon_\text{AlGaAs}/d_\text{AlGaAs}\times d_\text{PMMA}/\epsilon_\text{PMMA}\simeq 6.8$ in fair agreement with our first estimate. From the equation above it is evident, that a thicker insulator layer between the screen gate and the top gate with a smaller dielectric constant would increase the screening effect.} 
\scrap{For an alternative estimation of $\cs/\ct^0$ we determine the carrier density of the 2DES at grounded gates, $\vt=\vs=0$, based on Hall measurements to be $\ns^0\simeq1.45\times10^{11}\,\text {cm}^{-2}$. Using a plate capacitor model accounting for two dielectric layers of equal thickness, PMMA and AlGaAs, we then find the dielectric constant $\epsilon_\text{PMMA}\simeq2.0$ where we used the literature value $\epsilon_\text{AlGaAs}=12.7$ and the measured depletion voltage $\vtd=-1.48\,$V. Our plate capacitor model then predicts $\cs/\ct^0\simeq1+\epsilon_\text{AlGaAs}/d_\text{AlGaAs}\times d_\text{PMMA}/\epsilon_\text{PMMA}\simeq 6.8$ in fair agreement with our first estimate.}

\section{hall measurements: carrier density and mobility}

\begin{figure}[htb]
\begin{center}
\includegraphics[width=1\linewidth]{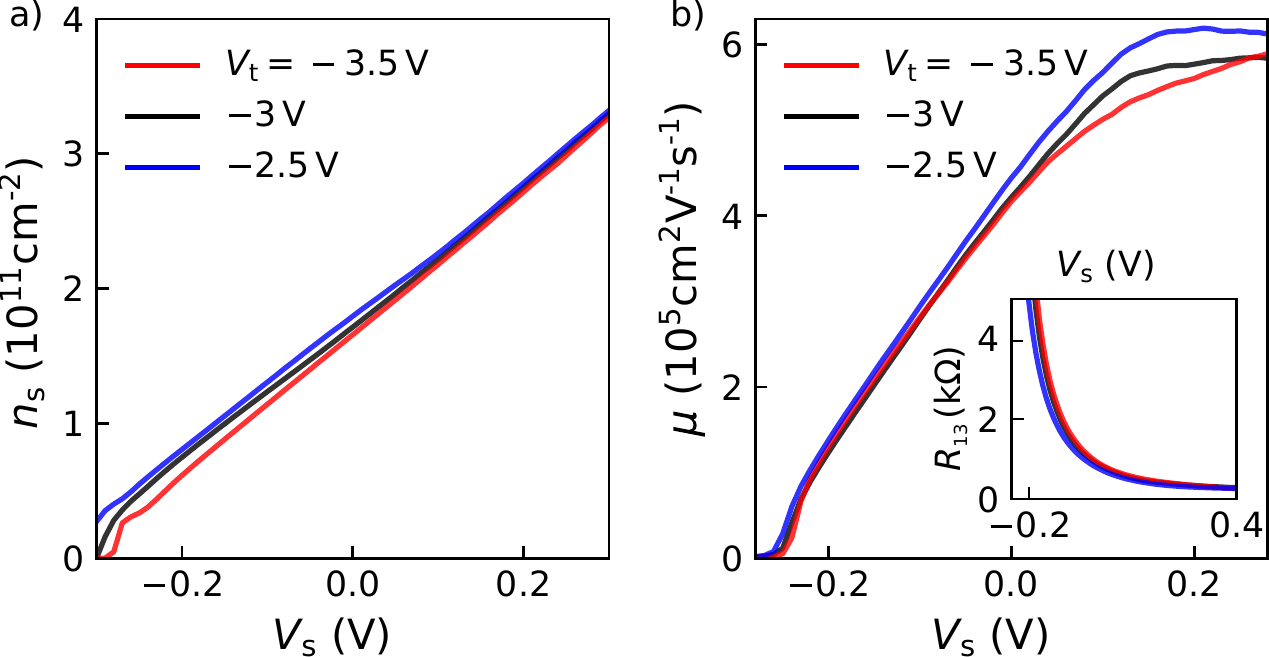}
\caption{Hall measurements: Electron carrier density $n_\text{s}$ in panel a) and mobility $\mu$ of the 2DES in panel b) versus screen gate voltage \vs\ for various top gate voltages $\vt<\vtd$. The inset presents the actually measured longitudinal resistance $R_{13}(\vs)$ at $B\to0\,$mT. }
\label{fig_3}
\end{center}
\end{figure}

We aim at evaluating the quality of the 2DES in nano circuits created with our method. Below we will use an Aharonov-Bohm ring for phase coherent measurements. However, first we measure carrier density and mobility based on the Hall-bar introduced above. As reference we use the ``nominal'' mobility and carrier density averaged over the wafer\SL{, which we}  measured directly after growth at the cryogenic temperature of $T=4,2\,$K \cite{wieck2000}. They are $\mu=0.7\times10^6\,\text{cm}^2$V$^{-1}$s$^{-1}$ and  $\ns=2.27\times10^{11}\,\text{cm}^{-2}$, corresponding to a mean free path of $l_\text m=5.5\,\mu$m. In our sample we determine the carrier density (averaged over the width of the Hall-bar) by measuring the classical Hall voltage $V_\text H\propto 1/\ns$ and the mobility by measuring the longitudinal resistance in the limit $B\to0$ ($R_{13}=R_{24}\propto\rho_0\propto (\ns\mu)^{-1}$), both at $T\simeq 4.2\,$K. In \fig{fig_3}{}
we present our results as a function of screen gate voltage \vs\ and for various top gate voltages $\vt<\vtd$, i.e.\ where the 2DES beyond the Hall-bar is fully depleted and the Hall-bar is well defined. Both, the carrier density and mobility depend only little on the top gate voltage but are widely tunable by varying the screen gate voltage. For $\vs<100\,$mV we observe a linear decrease of both, $\ns$ and $\mu$, with decreasing $\vs$ indicating an approximately constant capacitance \cs\ between 2DES and the screen gate and a resistivity $\rho_0\propto \ns^{-2}$ (equivalent to $\mu\propto \ns$). We note, that gate voltage independent capacitances (as our \cs) between gates and the 2DES are not guaranteed as this property depends on the wafer material.

In our sample, at $\vs=0$ carrier density and mobility are reduced by approximately a factor of two compared to the ``nominal'' values of the pristine wafer. However the ``nominal'' values can be recovered  by applying positive \vs. This result suggests, that wafers with a higher doping level could be advantageous for applications requiring a high mobility or a highly tunable carrier density.    

\section{Aharonov-Bohm measurements: phase coherence}

\begin{figure}[htb]
\begin{center}
\includegraphics[width=1\linewidth]{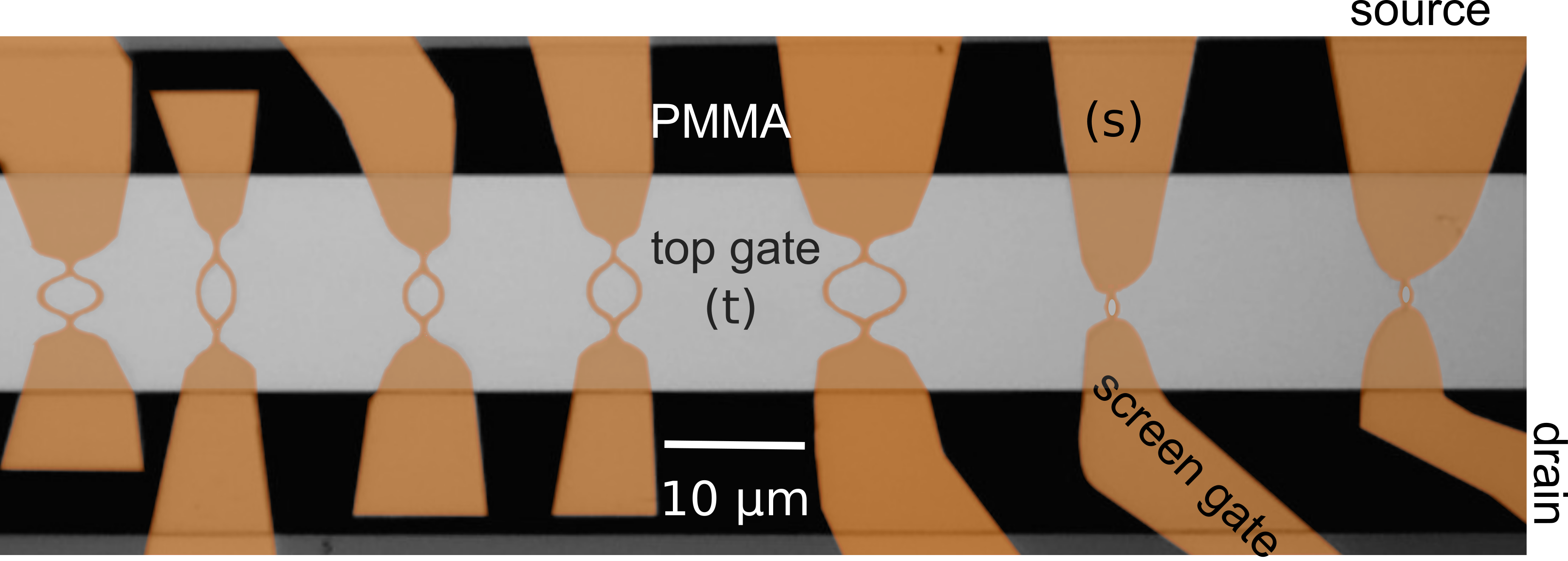}
\caption{False-colored optical microscope image of the Aharonov-Bohm sample with seven individual AB rings. The screen gates are colorized in orange and the top gate in gray. Unused AB rings are depleted by applying sufficiently negative \vs.}
\label{fig_4}  
\end{center}
\end{figure}

\begin{figure}[htb]
\begin{center}
\includegraphics[width=1\linewidth]{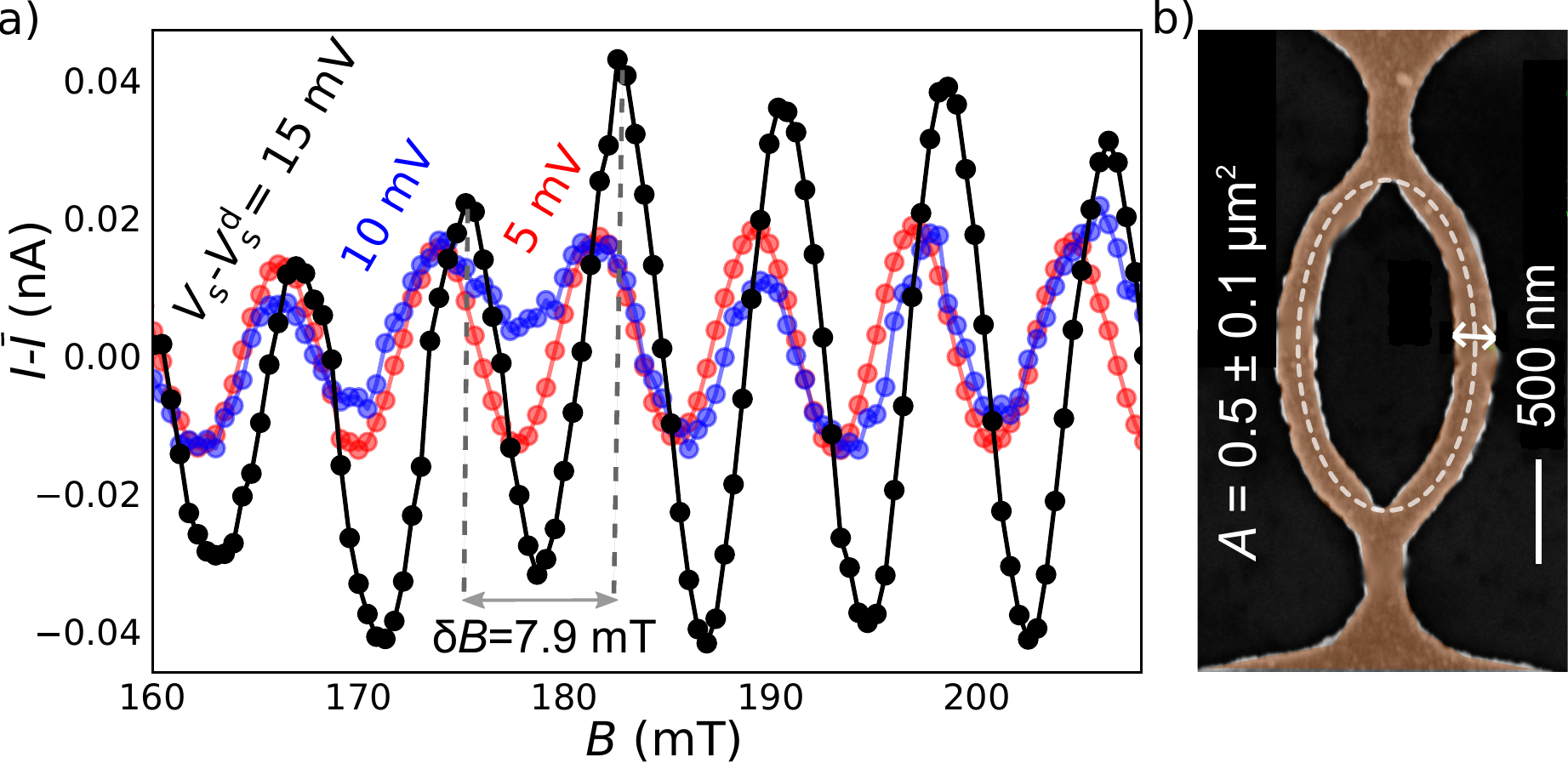}
\caption{a) Measured current $I-\overline I$ as a function of perpendicular magnetic field $B$ for three values of $\vs-\vs^\text d$ near depletion and $\vt=-3\,$V. (At $\vs^\text d$ the carriers beneath the screen gate are depleted.) The bath temperature is $T=25\,$mK and the source-drain voltage $V=0.1\,$mV.
b) Scanning electron microscope image of the measured AB-ring. The screen gate is shown in orange. The dashed white line embraces the area of $A=0.5\,\mu$m$^2$ corresponding to the measured magnetic field period of $\delta B=7.9\,$mT, see main text. The white double arrow indicates an error in $A$ of $\pm0.1\,\mu$m$^2$ corresponding to a maximum error in $\delta B$ of $\pm1.5\,$mT. This value reflects the experimental uncertainty in the tilt angle between the 2DES and the magnetic field of $\pm1^{\circ}$ ($B$ is the field component perpendicular to the 2DES).
}
\label{fig_5}
\end{center}
\end{figure}

Our method offers a straightforward way to fabricate conducting pathways with ring topology. In \fig{fig_4}{}
we present a photography of a sample containing seven quasi one-dimensional  Aharonov-Bohm (AB) rings of various sizes and shapes in a parallel configuration connected to two-dimensional leads. The conductance of an individual ring can be measured by depleting the 2DES below the top gate and below all ring-shaped screen gates besides the one of the AB-ring of interest.  To explore the phase coherence of the carriers, we here concentrate on the smallest ring (right most in \fig{fig_4}{}) which is also presented as a scanning-electron microscope picture in \fig{fig_5}{b}. 
In \fig{fig_5}{a} we present an example of AB oscillations, measured in a dilution refrigerator at a lattice temperature of 25\,mK. Plotted is the current $I$ flowing through our ring in response to a source-drain voltage of $V=0.1\,$mV versus the perpendicular magnetic field $B$. The AB oscillations can be formally described as
\begin{equation}
I=\overline{I}\left[1+v\cos\left(\frac{e}{\hbar}AB+\delta\varphi_\text{es}\right)\right]\,,
\label{eq_1}
\end{equation}
where $\overline{I}$ is the current averaged over $B$, $v=I_0/\overline I$ the visibility of the AB oscillations with amplitude $I_0$ and $A$ the area enclosed by the AB ring (which weakly depends on \vs).  The first term contained in the cosine is $2\pi$ times the number of enclosed magnetic flux quanta while $\delta\varphi_\text{es}$ sums up all other phase shifts which can be related to the existence of multiple paths (as for universal conductance fluctuations \cite{ginossar2010,ren2013,castellanos2013})  or geometry (such as the electrostatic AB effect \cite{wiel2003}). The measured period of the AB oscillation in \fig{fig_5}{a} of $\delta B\simeq7.9\,$mT corresponds to the enclosed area of $A=h/eB\simeq0.5\,\mu\text m^2$, coinciding with the area framed by the dashed line in \fig{fig_5}{b}.

In order to observe the AB-oscillations shown in \fig{fig_5}{a} it was necessary to almost completely deplete the carriers in the AB-ring by applying \vs\ close to the depletion voltage $\vs^\text d$.  This hints at a channel width so wide that it allows for multiple paths (in each arm) contributing with individual phases to the conductance which effectively reduces the visibility of the AB-oscillations \cite{giesbers2010}. As a rule of thumb, for our geometry an enclosed area difference of about 1\% would suffice to generate a phase shift of $\pi$ at $B\simeq200\,$mT. The almost depleted ring sufficiently reduces the number of possible paths to reach a visibility of few percent. Taking the Hall-bar measurements above as a reference for the applied gate voltages we expect a carrier density of $\sim8\times10^9\,$cm$^{-2}$ and a mean free path in the order of $1\,\mu$m which is the same order of magnitude as the arm length of our AB-ring of $L\simeq1.5\,\mu$m. However, screening is reduced along the almost depleted AB-ring, such that the mean-free-path could be shorter. Hence, we conclude that the electron dynamics in our AB-ring is located somewhere between the quasi-ballistic and the diffusive regime. One way to reach the ballistic regime in future devices will be to further reduce the intrinsic channel width such that quasi-one-dimensional channels can be realized at relatively large carrier densities. A further reduction of the screen gate width by a factor of four is easily achievable by electron-beam-lithography.

\begin{figure}[htb]
\begin{center}
\includegraphics[width=1\linewidth]{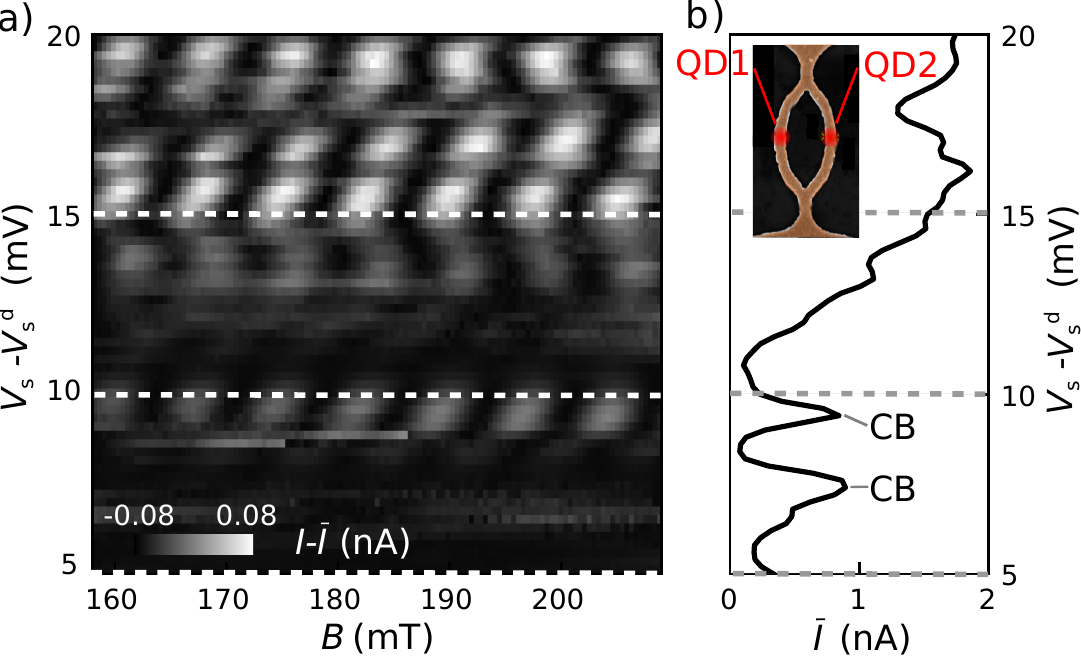}
\caption{a) Current oscillations $I-\overline{I}$ as a function of screen gate voltage and magnetic field at $V_\text{t}=-3\,$V. 
b) Coulomb blockade oscillations in $\overline I(\vs)$ (averaged over $B$). Two individual Coulomb blockade maxima are indicated with CB. The bath temperature was $T=25\,$mK and the source-drain voltage $V=0.1\,$mV. The data shown in Fig.\ \ref{fig_5}a are plots along the horizontal dashed lines.
}
\label{fig_52}
\end{center}
\end{figure}

In \fig{fig_52}a
we present AB-oscillations of the current $I-\overline I$ as a function of $B$ and \vs\ while in panel b we show an exemplary depletion curve plotting the current $\overline I(\vs)$  averaged over $B$. The latter strongly oscillates as observed for Coulomb blockade oscillations, the current becomes small but stays finite in the Coulomb valleys. Such a behavior can be explained assuming two quantum dots in parallel \cite{taylor1992}, i.e.\ one dot in each arm of the AB-ring as indicated in the inset. The exact position of the quantum dots is thereby unknown. The overall resistance of $R\simeq120\,\text k\Omega\gg h/e^2$ at the two distinct current maxima below $\vs-\vs^\text d=10\,$mV is in agreement with the assumption of two parallel dots giving rise to well established Coulomb blockade oscillations.

The two-terminal AB-oscillations in \fig{fig_52}a feature (i) continuous phase shifts at finite $B$, confirming the contribution of multiple paths in each arm, and (ii) phase jumps as a function of \vs, confirming the existence of quantum dots in the arms of the AB-ring (phase jumps have been previously observed for one dot in one arm) \cite{Avinun-Kalish2005,yacoby1996}. Note that our ring is too small to explain the observed phase jumps by means of the electrostatic AB-effect \cite{wiel2003}.
 
\begin{figure}[htb]
\begin{center}
\includegraphics[width=1\linewidth]{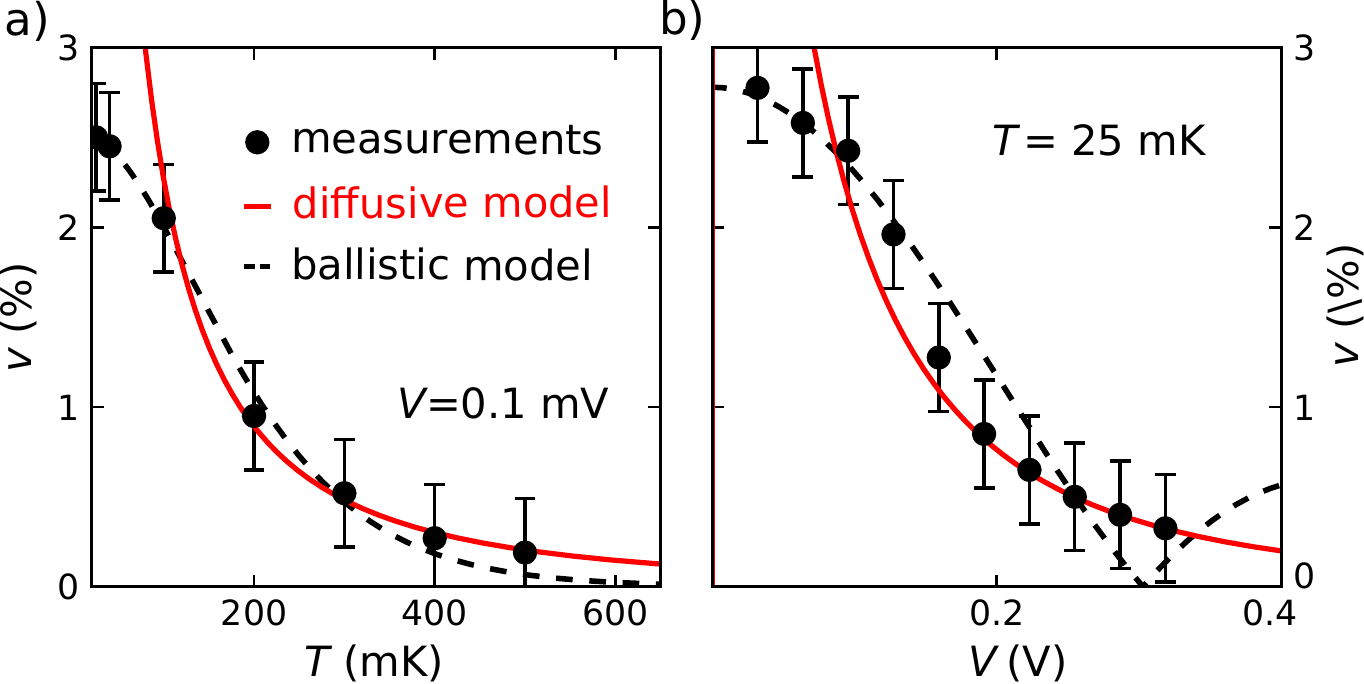}
\caption{Visibility $v(T)$ at $V=0.1\,$mV in panel a) and $v(V)$ at $T=25\,$mK in b); $\vs-\vs^\text d=15\,$mV and $\vt=-3\,$V.  Solid red lines are model curves assuming diffusive transport calculated with Eqs.\ (\ref{eq_diffusive}) in a) and (\ref{eq_diffusive2}) in b) for $v_0=56\%$, $E_\text{Th}=36\,\mu\text{eV}$, $\alpha=0.015\,\text{ps}\,\text{meV}^{2/3}$, $\beta=0.009\,\text{ps}\,\text{meV}^{2/3}$ and $\kappa=0.52$. Dashed black lines are calculated assuming ballistic transport with \eq{eq_ballistic} for $v_0=2.56\%$, $\Delta\tau=30\,$ps and $\kappa=0.52$.}
\label{fig_6}
\end{center}
\end{figure}

In the following we will discuss the dephasing as a function of temperature and source-drain voltage. In an ideal two-terminal AB-ring composed of one-dimensional arms dephasing by energy broadening is absent at modest energies. The reason is phase rigidity \cite{onsager1931,casimir1945,buttiker1988,yacoby1996,kreisbeck2010} allowing only phase shifts by multiples of $\pi$  which would require either very different arm length or an unreasonably large energy window. Such an ideal AB-ring would be a perfect device to study the electron-electron interaction \cite{washburn1985,russo2008,capron2013} remaining as possible dephasing process. However, realistic AB-rings as ours host multiple paths compromising the phase rigidity such that the temperature or source-drain voltage dependence of the dephasing at relatively small energies is dominated by energy broadening \cite{casse2000,buchholz2010,chung2005,litvin2007}. The measured temperature and source-drain voltage dependence of the visibility are presented in \fig{fig_6}{}
for $\vs-\vs^\text d=15\,$mV where the two-terminal resistance is $\simeq60\,\text k\Omega$, compare \fig{fig_52}b. Having already established the existence of two parallel quantum dots we now consider two scenarios, namely either ballistic or diffusive transport between the quantum dots. Searching for an answer we fit the measured data in \fig{fig_6}{} for two idealistic models. The first one assumes diffusive transport in an AB-ring with quasi-one-dimensional arms for which the temperature dependence of the visibility has been obtained from the weak localization theory \cite{washburn1986}
\begin{equation} \label{eq_diffusive}
v (T) =  v_{0} \left(\frac{E_\text{Th}}{k_\text{B}T}\right)^{1/2}\exp\left(\frac{-L}{\sqrt{D\tau_{\phi}}}\right)\,,
\end{equation} with $\tau_{\phi}=\alpha (k_\text{B} T)^{-2/3}$ \cite{treiber2009}.
This equation takes into account thermal broadening (square root term) and decoherence by scattering of electrons (exponential term). Here $E_\text{Th}=\hbar D/L^2$ is the Thouless energy, $D$ the 2D diffusion constant and $\tau_{\phi}$ the electrons decoherence time. The according voltage dependence of the visibility derived from non-equilibrium dephasing models is \cite{larkin1986,terrier2002}
	\begin{equation}
v(V) =  v_{0} \left(\frac{e\kappa V}{E_\text{Th}}\right)^{1/2}\exp\left(\frac{-L}{\sqrt{D\tau_{\phi}}}\right)
\label{eq_diffusive2}
\end{equation}
with $\tau_{\phi}=\beta (e\kappa V)^{-2/3}$ \cite{gougam2000,divincenzo1988}. The prefactor $\kappa=0.52$ takes into account that part of the source-drain voltage $V$ drops in the leads of the AB-ring. The red solid lines in \fig{fig_6}{} are fits to the respective temperature and voltage dependences given by \eq{eq_diffusive} and \eq{eq_diffusive2}. The diffusive model describes the measured data well for high energies but drastically overestimates the visibility at low $T$ or $V$. This deviation can be explained with the approximations  done in assuming $V=0$ for fitting the $T$-dependence and $T=0$ for fitting the $V$-dependence. The actual fit-parameters are listed in the caption of \fig{fig_6}{}.

In our second idealistic scenario we assume ballistic transport through the AB-ring. Because the dwell time $\simeq L/v_\text F$ of an electron moving ballistically through the AB-ring is short compared to $\tau_\phi$ in this case we can neglect the influence of Nyquist noise which leaves energy broadening as only remaining dephasing process \cite{treiber2009}. Combining voltage and temperature dependence in first order the ballistic scenario can be described by \cite{lin2010}:
\begin{equation} \label{eq_ballistic}
v= 2\pi v_{0}\frac{k_\text{B}T}{\left|e\kappa V\right|}\text{sinh$^{-1}$}\left(\frac{\pi k_\text{B}T}{\hbar/\Delta\tau}\right)\left|\sin\left(\frac{e\kappa V}{2\hbar/\Delta\tau}\right)\right|\,,
\end{equation}
where $\Delta\tau$ defines the difference of the dwell times of a ballistic electron in the two arms of the AB-ring.
A single fit to both data sets of \eq{eq_ballistic} representing the ballistic model is shown as black dashed lines in \fig{fig_6}{}. Our ballistic model describes the temperature dependence well but shows qualitative deviations in the voltage dependence (at high voltages). The actual fit-parameters are listed in the caption of \fig{fig_6}{}. We find a dwell time difference of $\Delta\tau=30\,$ps. On the one hand this corresponds to an unrealistically large arm length difference of $\sim1\,\mu$m assuming ballistic motion at the Fermi velocity. On the other hand, the existence of a quantum dot in each arm leads to multiple reflections which would enhance dwell times. As a result, without further experimental and theoretical efforts it is impossible to determine from our data, whether transport through the AB ring is  diffusive or ballistic. Diffusive transport might be caused by the almost complete depletion in the AB-rings which is necessary to reduce the number of one-dimensional channels preventing a higher visibility. We believe that AB-rings with narrower arms but higher carrier density will in future help to reach ballistic transport and to reduce the chance of the formation of quantum dots.
 
\section{conclusion}

We have explored an alternative method to define mesoscopic circuits in heterostructures based on the electric field effect. The idea is to deplete most of the 2DES by means of a global top gate. Only at those areas where carriers are needed screen gates placed below the top gate are used to shield the effect of the top gate locally. The resulting circuits are highly tunable on the nanoscale as demonstrated in the presented experiments. Importantly, our method has the advantage of reducing the complexity of gate defined nanostructures. In more detail, it allows a straightforward way to realize conducting paths with ring topology and offers a way to define complex structures with a smaller number of gates compared to the conventional technology based on multiple depletion gates. Our Aharonov-Bohm measurements demonstrate phase coherence comparable to that in conventional AB-rings in semiconductors which makes our method suitable for quantum information applications. While not demonstrated here the closer vicinity of a metal gate to the carriers is expected to lead to a reduction of the Coulomb interaction between carriers. As such our method can be viewed as an alternative which will allow to increase the variety of physical properties in nanocircuits. Future tasks will include the definition of quantum point contacts and chains of quantum dots by using screen gates.
\\
\section{Acknowledgments}

We thank P.\ Brouwer for fruitful discussions and are grateful for financial support from the DFG via LU 819/4-1.

\nocite{apsrev41Control}
\bibliographystyle{apsrev4-1}
\bibliography{literature}

\end{document}